\documentclass{article}
\usepackage{waspaa23,amsmath,graphicx,url,times}
\usepackage{amsfonts}
\usepackage{color}
\usepackage{hyperref}
\usepackage{soul}
\usepackage{bbm}
\usepackage{multirow}

\title{AUDIO INPUTS FOR ACTIVE SPEAKER DETECTION AND LOCALIZATION VIA MICROPHONE ARRAY}

\name{Author 1,$^{1}$
      Author 2,$^{2}$
      Author 3,$^{3}$
      Author 4$^{2}$}
\address{$^1$ Author names and affiliations omitted for double-blind review.\\
         $^2$ Please do not make changes to this section.\\
         $^3$ Author information may be added after paper acceptance.\\
}

\name{Davide Berghi, Philip J. B. Jackson}
\address{Centre for Vision, Speech and Signal Processing, University of Surrey, Guildford, UK.}

\begin{document}

\ninept
\maketitle

\begin{sloppy}

\begin{abstract}
This study considers the problem of detecting and locating an active talker's horizontal position 
from multichannel audio captured by a microphone array. 
We refer to this as active speaker detection and localization (ASDL). 
Our goal was to investigate the performance of spatial acoustic features extracted from the multichannel audio as the input of a  convolutional recurrent neural network (CRNN), in relation to the number of channels employed and additive noise. 
To this end, experiments were conducted to compare the generalized cross-correlation with phase transform (GCC-PHAT), the spatial cue-augmented log-spectrogram (SALSA) features, and a recently-proposed beamforming method, evaluating their robustness to various noise intensities. 
The array aperture and sampling density were tested by
taking subsets from the 16-microphone array. 
Results and tests of statistical significance demonstrate the microphones' contribution to performance on the TragicTalkers dataset, which offers opportunities to investigate audio-visual approaches in the future.

\end{abstract}

\begin{keywords}
Features extraction, active speaker detection and localization, microphone array, multichannel audio
\end{keywords}

\section{Introduction}
\label{sec:intro}

Active speaker detection and localization (ASDL) incorporates two subtasks: detecting the presence of speech and regressing the position of the speaker in the video frames. This task is vital in applications such as speaker diarization \cite{Gebru:2018:SpeakerDiarization}, human-robot interaction \cite{He:2018:deepNetsHRI}, or VR/AR \cite{jiang2022egocentric}.
It can be tackled in two separate stages. The first corresponds to the localization subtask and is performed by a visual face detector that selects a set of hypothetical candidate speakers. 
The second classifies the detected faces as active/inactive. This classification stage is referred to as active speaker detection (ASD) and is typically performed as an audio-visual problem with single-channel audio \cite{Roth:2020:AVA,Chung2019NaverAA,Alcazar_2020_CVPR,zhang2021unicon,tao2021TalkNet}.
In contrast, we investigate ASDL using only the audio modality with multichannel signals to simultaneously detect and locate the speaker. 
The advantage of employing multichannel audio instead of a visual face detector for the regression subtask is that it does not fail when the face of the speaker is not visible from the camera point of view, e.g., with visual occlusions.
Under this formulation, ASDL can be thought of as a specialization of sound event localization and detection (SELD) \cite{Adavanne:2019:SELDnet} with speech as the only class. SELD has been widely explored in recent years \cite{Nguyen:2020:SMN,Shimada:2021:ACCDOA,cao:2020:EIN,cao:2021:EINv2}, partially thanks to the introduction of the SELD challenge in the DCASE Task3 \cite{Politis2022DCASE}.
Research has focussed on two spatial formats of the audio provided: first-order ambisonics (FOA) or microphone array channels (MIC).
That community's input features, in particular for the latter format, inspired this study. 
For simplicity, we addressed the horizontal ASDL problem aiming to focus our study on the performance of spatial acoustic feature sets extracted from the multichannel audio captured by a microphone array. The data utilized in this study was recorded with a co-located audio-visual sensing platform comprised of multiple cameras and a microphone array. We leverage only the audio data to regress the speaker's position directly in the horizontal domain of each camera.

We evaluate the popular generalized cross-correlation with phase transform (GCC-PHAT) \cite{Knapp:gccphat:1976,Cao:2019:polyphonic}, the spatial cue-augmented log-spectrogram (SALSA) features \cite{Nguyen:2021:SALSA,Nguyen:2021:SALSALiteAF}, and a beamforming approach \cite{Berghi:2021:mmsp} on the 16-channel audio data of the TragicTalkers dataset \cite{Berghi:CVMP:2022}, which has audio-visual speech recordings with multiple cameras and microphones simultaneously in a sound-treated research lab.
Experiments include varying the number of input microphones to test various subset combinations and array apertures, as well as the robustness of the input features to additive noise corruptions.
For fairness, all the input features are compared utilizing the same network architecture back-end. We adopted the CRNN as a capable yet classic form of deep neural network, as in many recent audio AI-based works \cite{Adavanne:2019:SELDnet,Cao:2019:polyphonic,Nguyen:2021:SALSA,Nguyen:2021:SALSALiteAF}.
To the best of our knowledge, we are the first to provide a feature experimental comparison in such conditions on a large 16-element microphone array. 

\section{Spatial Input Features} 
\label{sec:featuresSELD}

Two popular types of input features are tested in this study: the GCC-PHAT \cite{Cao:2019:polyphonic} and the SALSA-Lite features \cite{Nguyen:2021:SALSALiteAF}. In addition, a beamformer-based method 
is included.  

{\bf GCC-PHAT.}
GCC-PHAT is employed to estimate the time-difference-of-arrival (TDOA) of a sound source at two microphones \cite{Knapp:gccphat:1976}. 
The GCC-PHAT between the \textit{i}-th and the \textit{j}-th microphone is defined at each audio frame \textit{t} as:
\begin{equation}
    \label{gcc_phat_eq}
    GCC_{ij}(t,\tau)=\mathcal{F}^{-1}_{f\rightarrow\tau} \frac{\mathbf{X}_i(t,f)\mathbf{X}^*_j(t,f)}{|\mathbf{X}_i(t,f)||\mathbf{X}^*_j(t,f)|},
\end{equation}
where $\mathbf{X}_{i}(t,f)$ is the Short-Time Fourier Transform (STFT) of the \textit{i}-th channel, $\mathcal{F}^{-1}_{f\rightarrow\tau}$ the inverse-FFT from the frequency domain $f$ to the lag-time domain $\tau$, and $(.)^*$ denotes the complex conjugate. The TDOA can be estimated as the lag-time $\Delta\tau$ take maximizes $GCC_{ij}(t,\tau)$.
Representing it as a `spectrogram' with time-lags $\tau$ on the frequency axis 
allows concatenation with the log-mel spectrograms of the microphone array's channels, as indicated by Cao \textit{et al.} \cite{Cao:2019:polyphonic}. 
The maximum number of delayed samples corresponding to $\Delta\tau_{max}$ is computed as $d_{max}/c\cdot{f_s}$, where $c$ is the speed of sound, $f_s$ the sampling frequency and $d_{max}$ the maximum distance, or aperture, between the two furthest microphones. The number of mel-frequency bins must be greater or equal to $2\cdot{\Delta\tau_{max}}+1$ considering delay and advance between the signals \cite{Cao:2019:polyphonic}. 
This study only analyzes the frontal horizontal domain as the speakers are always contained in the camera field of view (FoV) so, the formula can be modified as: $\Delta\tau_{max}=d_{rel}/c\cdot{f_s}$, with $d_{rel}$ representing the relative maximum distance between the microphones, given by $d_{max}\cdot{sin{\tfrac{\theta}{2}}}$, where $\theta$ is the camera horizontal FoV in degrees. 
Typically the GCC-PHAT features are computed for each possible microphone pair of the array (i.e. $M(M-1)/2$ pairs for an $M$-element microphone array) and concatenated with the log-mel spectrograms of each audio channel. Here, since all microphones face forwards creating a planar array, a single log-mel spectrogram from one microphone is employed to reduce the input dimensionality. 
To further contain the number of microphone combinations and keep the input small, a reference-based (ref.) approach is also proposed where the GCC-PHAT is only computed between a reference microphone and the other channels, i.e., $M-1$ pairs, instead of all possible pairs. 
Hereafter, the term GCC-PHAT denotes the concatenation of GCC-PHAT features with a single log-mel spectrogram.

\begin{table}[tb]
\caption{Summary of the input features employed.}
\begin{center}
\begin{tabular}{p{0.21\columnwidth}|p{0.39\columnwidth}|p{0.24\columnwidth}}
\hline
\textbf{Feature set}&\textbf{Components}&\textbf{\# channels} \\
\hline
Beamformer & Steered log-mel spect. & \# look dir. \\
\hline
GCC-PHAT & 1 log-mel spect. + 

GCC-PHAT `spect.' & 1+$M$($M$-1)$/$2 \\
\hline
GCC-PHAT (ref.) & 1 log-mel spect. + 

GCC-PHAT `spect.' & 1+($M$-1) \\
\hline
SALSA-IPD & 1 log-lin spect. + IPD & 1+($M$-1) \\
\hline 
SALSA-Lite & 1 log-lin spect. + NIPD & 1+($M$-1) \\
\hline

\end{tabular}\vspace{-3ex}
\label{tab_inputsDescript}
\end{center}
\vspace{-3ex}
\end{table}

{\bf SALSA.}
Nguyen \textit{et al.} proposed SALSA-Lite \cite{Nguyen:2021:SALSALiteAF}, a lighter variation of SALSA \cite{Nguyen:2021:SALSA}: instead of eigenvector-based phase vector (EPV), it uses a normalized version of the inter-channel phase difference (NIPD), computed for each time-frequency (TF) bin, concatenated with multichannel log-linear spectrograms. 
Preliminary tests on the dataset found that SALSA-Lite performs on par with SALSA while being remarkably faster to extract. Therefore, SALSA-Lite has been adopted in this study. 
NIPD $\in\mathbb{R}^{M-1}$ is computed with respect to a reference microphone.

A variant of SALSA-Lite is SALSA-IPD, which computes the inter-channel phase difference (IPD) but does not apply frequency normalization. Thus, SALSA-IPD uses frequency-dependent IPD, normalized by $(-2\pi)$, 
instead of $(-2\pi f/c)$. 
Both SALSA-Lite and SALSA-IPD are tested.
Thanks to the extraction of directional cues at each TF bin, SALSA-based input features compactly
align log-linear spectrograms with NIPD.
As per GCC-PHAT, we appended a single log-linear spectrogram in the SALSA feature sets.

{\bf Beamformer.}
\label{subsec:BF}
We adopted a beamforming-based approach for ASDL feature extraction as described in \cite{Berghi:2021:mmsp}.
The approach steers the array's listening direction toward a set of horizontal `look' directions by spatial filtering.
Using Galindo \textit{et al.}'s beamforming toolbox \cite{galindo:2020:microphone}, a set of super-directive beamformer (SDB) weights was computed to spatially filter the array signals over 15 frontal look directions 
to yield one audio signal for each direction. 
These are converted to 15 steered log-mel spectrograms  and concatenated to form the input. 
The look directions are equally spaced at 5° around the horizon in the range $\pm$30°, plus two wider directions at $\pm$45°. 
A summary of the input feature sets adopted in this study is proposed in Tab.~\ref{tab_inputsDescript}, including the number of channels generated by each method for an $M$-element array.


\section{Method}
\label{sec:met}


\textbf{Dataset.} Tests are conducted on the TragicTalkers dataset \cite{Berghi:CVMP:2022}, which was captured with audio-visual arrays (AVA) rigs: multi-sensing platforms each comprising 11 cameras and a 16-element planar microphone array (see Fig.~\ref{fig:AVArig}\,(b)). 
The dataset includes two actors, whose speech does not overlap. The distance between the actors and the microphone array is in the range of 3-4 m and the studio reverberation time is 0.3s in the mid 0.5-2kHz range. 
The multi-view video allows extension of the audio network's training via an additional task: choosing the view for speaker localization. 
A one-hot vector denoting the selected view is appended to the audio input, providing data augmentation through camera perspective variations. 
TragicTalkers is comprised of about 3.8h of video data: 3.1h used for training and 0.7h for testing.
Therefore, the network learns the correct mapping from the input audio features to the desired camera view.
To supervise the training of the audio network, pseudo-labels were automatically generated using a face detector and a pre-trained audio-visual ASD model, as described in \cite{Berghi:2021:mmsp}.

\begin{figure}[tb]
\centerline{\includegraphics[width=\columnwidth]{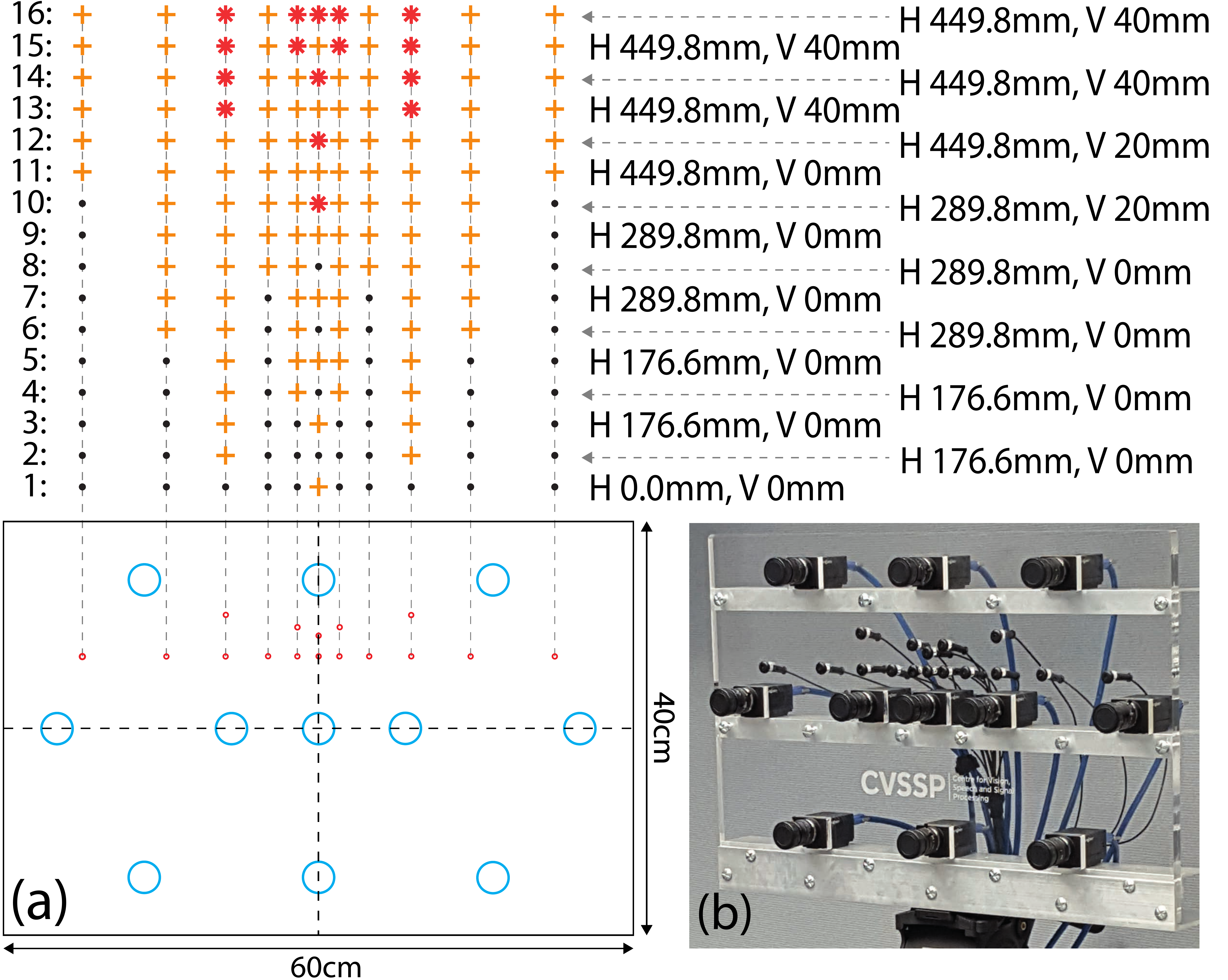}}
\caption{(a) Schematic of camera (blue circles) and microphone (red dots) positions on the AVA rig. (b) Photo of an AVA rig. Above the schematic are the microphone combinations used in the experiments with varying number of microphones, from 1 (mono) to 16 (full house). Orange crosses: 1 selected; Red asterisks: 2 selected; Dots: unused; where the mics are always taken from the lower line subarray first and the upper group of mics second. `H' and `V' indicate the horizontal and vertical array aperture, respectively.}
\label{fig:AVArig}
\vspace{-3ex}
\end{figure}

\textbf{Network Architecture.} The CRNN architecture (Tab.\,\ref{tab_architecture}) takes in input the audio features with shape $C_{in}\times T_{in}\times F_{in}$, as the number of channels, temporal bins, and frequency bins, respectively. 
$C_{in}$ depends on the feature set. 
Inspired by \cite{Cao:2019:polyphonic}, the network's convolutional backbone 
reduced the temporal and frequency resolution by a factor of 16.
Frequency-average pooling is then applied and the tensor reshaped to $T_{in}/16\times 512$. 
We employed two bidirectional gated recurrent units with 256 hidden units, preserving the output shape. 
Finally, the tensor is fed to two fully-connected layers that reduce the feature map's output to 2 values per time frame: a regression value $x_i$ and respective speech activity confidence $C_i$ for each of the $T_{out}$ time frames, normalized in the range [0, 1] by Sigmoid function. 
$T_{in}$ is chosen so that $T_{in}/16=T_{out}$ matches the label frame rate.  
Between the two FC layers, each feature map is concatenated with an 11-dimensional one-hot vector encoding the camera view for regression (CamID). 

\textbf{Training.}
Our audio CRNN is trained on the horizontal positions of the active speakers in the video frames with pseudo-labels from a pre-trained audio-visual ASD model \cite{Alcazar_2020_CVPR}.
To train the network, a sum-squared error-based loss computed at each output frame prediction was employed.
The loss is composed of a regression loss and a confidence loss:
\begin{equation}
Loss=\sum_{i=1}^{T_{in}/16}\mathbbm{1}_i(x_i-\hat{x}_i)^2+(C_i-\hat{C}_i)^2
\label{loss_function}
\end{equation}
where $x_i$ and $\hat{x}_i$ are respectively the speaker's predicted and target positions (pseudo-labels) along the horizontal axis in the $i$-th video frame, normalized in the range [0, 1], while $C_i$ and $\hat{C}_i$ are predicted and target confidences. 
$\hat{C}_i$ is 1 for active frames, 0 for silent. 
The masking term $\mathbbm{1}_i$ is 1 only when the frame is active and the speaker is detected by the ASD (the pseudo-label is available); 0 otherwise.

\section{EXPERIMENTS}
\label{sec:exp}

\subsection{Implementation Details and Evaluation Metrics}

Positive pseudo-labels are manually screened to remove potential false positives in the training dataset. The network is trained with a 5-fold cross-validation approach.  Evaluation is performed on the TragicTalkers test set, labeled for speaker mouth positions.

Audio clips (48\,kHz sample frequency) are extracted 2\,s long 
with 1\,s overlap for training.
An STFT with 512-point Hann window and hop size of 100 
generated spectrograms discretized into 960 temporal bins ($T_{in}$), 
providing resolution ($T_{in}/16$) to match the labels' video frame rate (30fps). 
For the log-mel spectrograms, 64 mel-frequency bins are used. Instead, for the log-linear spectrogram and the IPD/NIPD used in SALSA-based features, an upper cutoff frequency of 6\,kHz is applied to only extract the first 64 frequency bins. 
The maximum distance $d_{max}$ between two microphones is 450\,mm and the camera's horizontal FoV is 55°. Therefore, 64 time-lags are enough with GCC-PHAT for speakers in view, with the single-channel log-mel spectrogram appended. 
The network is trained for 50 epochs using batches of 32 inputs and Adam optimizer. 
The learning rate is fixed for the first 30 epochs, then reduced by 10\% every epoch.
The first microphone from the lower subarray was used as the reference for GCC-PHAT and SALSA, as tests on the central microphone gave a poorer performance.

\begin{table}[tb]
\caption{CRNN Network Architecture.}
\begin{center}

\begin{tabular}{c|c}
\hline
\textbf{Layer}&\textbf{Output size} \\
\hline
Input features & $C_{in}\times T_{in}\times F_{in}$ \\
\hline
\multirow{2}{*}{\LARGE{((}}  Conv. 3$\times$3  \multirow{2}{*}{\LARGE{)}}\multirow{2}{*}{$\times2$} \multirow{2}{*}{\LARGE{)}}\multirow{2}{*}{, AvgPool}\multirow{2}{*}{\LARGE{)}}\multirow{2}{*}{$\times4$} & \multirow{2}{*}{512 $\times\dfrac{T_{in}}{16}\times\dfrac{F_{in}}{16}$} \\

BN, ReLu\hspace{24 mm}  & \\

\hline
Freq. AvgPool , Reshape & $T_{in}/16\times 512$  \\
\hline
biGRU $\times2$ & $T_{in}/16\times 512$ \\
\hline 
FC1 & $T_{in}/16\times 32$ \\
\hline
Concat. CamID & $T_{in}/16\times 43$ \\
\hline
FC2 & $T_{out}\times 2$ \\
\hline

\end{tabular}\vspace{-3ex}
\label{tab_architecture}
\end{center}
\vspace{-3ex}
\end{table}



A frame prediction is positive 
when the predicted confidence is above a threshold, and a positive detection is true when the localization error is within a predefined spatial tolerance. 
The precision and recall rates are computed by varying the confidence threshold from 0\% to 100\% sampling the thresholds from a Sigmoid-spaced distribution to provide more data points for high and low confidence values. 
The average precision (AP) was computed as the numerical integration of the precision-recall curve, as indicated in \cite{Everingham:2015:pacalVOC}. 
We set a spatial tolerance of $\pm$2° along the azimuth according to human auditory perception \cite{Strybel:2000:MinimumAA}, 
corresponding to $\pm$89 pixels on the image plane. 
From the precision and recall rates, the F1 score is computed too. 
The average distance (aD) and the detection error (Det Err \%) are the metrics employed to evaluate the localization and the speech detection subtasks, respectively. 
We set a 0.5 confidence threshold to binarize the active-silent predictions in the detection subtask.
To run tests of statistical significance and standard errors for methods' comparison, the metrics were also separately computed for each test sequence in order to generate multiple data points.  

\subsection{Methods and Baselines}

In addition to the features of interest described, two reference methods that do not perform spatial processing are provided. 
\\
\textbf{LogMelSpec\,-\,16mics:} Network trained using the log-mel spectrograms of the 16 microphones directly. Neglecting the phase component has been explored in past audio-visual research \cite{Chen2020SemanticAN,Chen:2021:LearningWaypoints,Gan2019SelfSupervisedMV}  
\\
\textbf{MagPhSpec\,-\,16mics:} Network trained with the concatenation of magnitude and phase spectrograms extracted from the 16 microphone signal, as in SELDnet \cite{Adavanne:2019:SELDnet}.
\\
\textbf{BF\,-\,\textit{n} dir:} Network trained using the beamformer-based audio features. In addition to \textit{n}$=$15 look directions as described previously, a fewer number of directions is tested too. `BF\,-\,3 look dir' employs the directions at 0° and $\pm$20°; while `BF\,-\,7 look dir' the directions at 0°, $\pm$15°, $\pm$30° and $\pm$45°.  \\
\textbf{GCC-PHAT\,-\,\textit{M}} mics: GCC-PHAT audio feature extractor applied to \textit{M} microphones. When specified by `(ref.)', the GCC-PHAT is computed with respect to the first microphone only. If not specified, all possible microphone pairs are used. The number of employed microphones \textit{M} is varied between 2 and 16 and the subset of microphones for each value of \textit{M} is selected as depicted in Fig.\,\ref{fig:AVArig} (a).
\\
\textbf{SALSA-Lite\,/\,SALSA-IPD\,-\,\textit{M}} mics: SALSA-based features computed with respect to the first microphone. The \textit{M} microphones are selected as in the `GCC-PHAT\,-\,\textit{M} mics' method.

\subsection{Experimental Results}

Tab.~\ref{tab:Baselines} shows the results achieved with the three feature sets and the two reference methods. 
`MagPhSpec\,-\,16mics' performs better than the plain log-mel spectrograms thanks to the directional cues encoded in the phase. However, all three types of feature extractors provide a great benefit compared to the reference methods due to the extraction of important spatial information.
A t-test analysis on the F1 scores achieved with `SALSA-Lite\,-\,16mics' and `BF\,-\,15 dir' statistically supports the superiority of the former (p=0.015).
Although the beamforming-based features achieve performances weaker than GCC-PHAT or SALSA, their dimensionality does not depend on the number of microphones. In particular, the GCC-PHAT computed on all microphone pairs generates extremely high-dimensional inputs when many microphones are employed. 

The adoption of the GCC-PHAT or SALSA-based features enables great performances with just two microphones: `GCC-PHAT\,-\,2mics' reaches an F1@2° of 0.82, `SALSA-IPD\,-\,2mics' and `SALSA-Lite\,-\,2mics' 0.85 and 0.86, respectively.
When additional microphones are included, the performance further increases as shown in Fig.~\ref{fig:F1_over_mics}. 
Overall, the F1 score increases when more microphones are employed. However, the performance is mainly influenced by the aperture of the array. When \textit{M} is between 2 and 5 microphones, the array's horizontal aperture is fixed to 177\,mm and the performance is quite stable. 
As \textit{M} is increased to 6 microphones, the horizontal aperture is also increased to 290\,mm and a beneficial effect in all four methods is produced. The positive trend persists for $M$\,$=$\,7. 
However, it seems to stabilize as more microphones are added, even decreasing with 9 and 10 microphones. For $M$\,$\geq$\,11 the full horizontal aperture is employed (450\,mm), and the trend increases again ($M$\,$=$\,11, $M$\,$=$\,12) but it then tends to remain stable when additional microphones are included.
This behavior suggests that most of the gain is given by the aperture, while a denser sampling at several intermediate spatial positions does not always produce a beneficial effect, adding complexity to the input tensor. 
In general, SALSA-based features achieve the greatest performance. This is probably due to the compactness of their representation that offers one-to-one pixel mapping between the spectrogram and the NIPD/IPD representations. 
It is preferable to employ the GCC-PHAT with a single reference microphone as it often provides better results while being easier to handle and faster to compute.

\begin{figure}[tb]
\centerline{\includegraphics[width=\columnwidth]{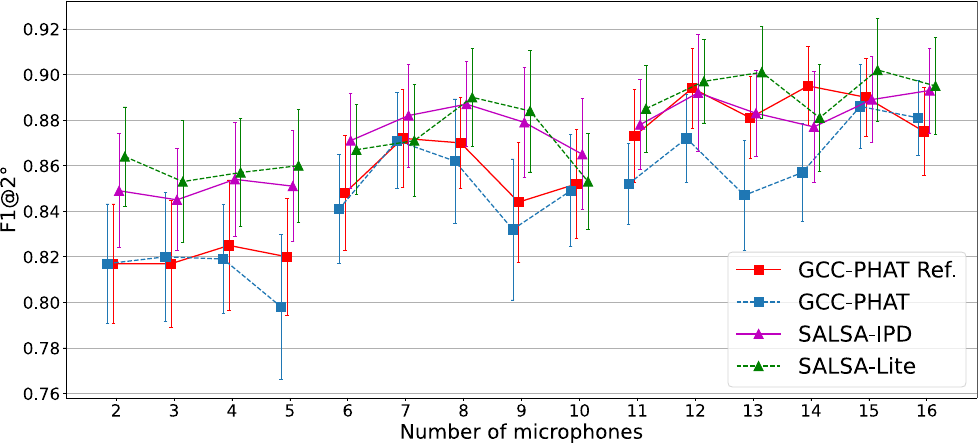}}
\caption{F1@2° vs.\ microphone count for 4 feature sets. Results are grouped by array apertures with slight offsets around the microphone integer to facilitate visualization.}
\label{fig:F1_over_mics}
\vspace{-3ex}
\end{figure}

\begin{table}[tb]
\caption{Features' detection error (Det Err), average distance (aD), average precision (AP) and F1 at 2°. Results are with \textit{M}\,=\,16 mics.} 
\begin{center}
\begin{tabular}{c|c|c|c|c|c}
\hline
\textbf{Features}&$\mathbf{C_{in}}$&\textbf{DetErr}&\textbf{aD}&\textbf{AP}&\textbf{F1} \\
\hline
LogMelSpec & 16 & \textbf{3.3\%} & 85p & 44.7\% & 0.621 \\
MagPhSpec & 32 & 3.5\% & 58p & 66.8\% & 0.777 \\
\hline
BF-3 dir & 3 & 3.7\% & 53p & 73.4\% & 0.819 \\
BF-7 dir & 7 & 3.9\% & 50p & 74.8\% & 0.834 \\
BF-15 dir & 15 & 3.6\% & 46p & 77.1\% & 0.850 \\
\hline 
GCC-PHAT\,{(ref)} & 16 & 4.0\% & 42p & 82.4\% & 0.875 \\
GCC-PHAT & 121$^{*}$ & 4.0\% & 41p & 83.3\% & 0.881 \\
\hline
SALSA-IPD & 16 & 3.8\% & \textbf{39p} & 85.4\% & 0.893 \\
SALSA-Lite & 16 & 4.3\% & \textbf{39p} & \textbf{86.1\%} & \textbf{0.895} \\
\hline
\end{tabular}\vspace{-3ex}
\label{tab:Baselines}
\end{center}
\vspace{0.8ex}
\footnotesize{$^{*}$Information loss may arise as $C_{in}$ is greater than the number of channels after the first convolutional block.}
\vspace{-3ex}
\end{table}

To further investigate the effect of the array's aperture across different feature sets, statistical significance is tested between the results of SALSA-Lite and GCC-PHAT (ref). 
Three groups of microphones are investigated according to the three array's apertures, i.e., from 2 to 5 microphones (177\,mm), from 6 to 10 microphones (290\,mm), and from 11 to 16 microphones (450\,mm). 
With the first and second groups, SALSA-Lite proved to be significantly better than GCC-PHAT (ref) (p=3E-7 and p=5E-3); while the third group gave no statistical significance. 
This suggests that SALSA-Lite is preferable with fewer microphones and compact arrangements.

To compare the robustness of the GCC-PHAT (ref) and SALSA-Lite features, we re-trained the network corrupting the dataset with additive pink noise at SNRs from 0 dB to 40 dB for the 16 and 4 microphone cases. 
F1 scores are reported in Tab.\,\ref{tab:SNR}. Note that the performance for the clean cases improves compared to the previous results achieved by training the network only on clean data as this provides a form of data augmentation, increasing the size of the training set.
With 16 mics, both feature types achieve similar performance for all noise intensities, with GCC-PHAT (ref) slightly outperforming SALSA-Lite but by less than 1\% and without statistical significance. 
Interestingly, for the 4 mics case, GCC-PHAT (ref) performs roughly on par with SALSA-Lite on the benign clean and 40 dB sets, contradicting our earlier observation where SALSA-Lite was stronger with fewer microphones. This suggests that GCC-PHAT (ref) might require more training data to perform well
on compact arrays. 
However, as the noise intensity increases, SALSA-Lite appears remarkably robust in the 4mics case, exceeding GCC-PHAT (ref)'s performance by 4\% at the lowest SNRs.

\begin{table}[tb]
\caption{F1 score on data corrupted with additive noise}
\begin{center}
\begin{tabular}{c|c|c|c|c}
\cline{2-5}
& \multicolumn{2}{c|}{\textbf{GCC-PHAT (ref)}} & \multicolumn{2}{c}{\textbf{SALSA-Lite}} \\
\hline
\textbf{SNR}& \textbf{16mics} & \textbf{4mics} & \textbf{16mics} & \textbf{4mics} \\
\hline
\textbf{Clean} & {0.928} & 0.887 & 0.919 & 0.881 \\
\hline
\textbf{40 dB} & {0.927} & 0.883 & 0.918 & 0.882 \\
\hline
\textbf{30 dB} & {0.925} & 0.874 & 0.921 & 0.880 \\
\hline
\textbf{20 dB} & {0.913} & 0.846 & 0.911 & 0.858 \\
\hline
\textbf{10 dB} & {0.875} & 0.747 & 0.866 & 0.787 \\
\hline
\textbf{0 dB} & {0.740} & 0.580 & 0.735 & 0.624 \\
\hline

\end{tabular}\vspace{-3ex}
\label{tab:SNR}
\end{center}

\vspace{-2ex}
\end{table}

\section{CONCLUSION}

This paper compares several types of audio input features addressing the ASDL task on the TragicTalkers dataset. 
We tested SALSA-based features and the GCC-PHAT computed over all possible microphone pairs or with a reference microphone. 
Beamformer-based input features were studied too.
While the reference LogMelSpec method had up to 1\% lower detection error, the GCC-PHAT and SALSA features halved the localization error, yielding a substantial increase in the F1 score from 0.6 to 0.9. 
A comprehensive study of the effects of a varying number of microphones on the overall performance was conducted. Results suggest that performance tends to increase with the number of microphones but seems mostly influenced by the aperture of the array. 
SALSA-Lite performs significantly better than GCC-PHAT with fewer microphones. However, augmenting the audio with additive noise eliminates this disparity. Therefore, we hypothesize that GCC-PHAT requires larger datasets to perform on par with SALSA-Lite. 
Nevertheless, SALSA-Lite with compact array configurations is more robust under heavy noise conditions.
In the 16mics case, GCC-PHAT with a reference microphone appears equally robust.   
Further research may explore the optimal combination of sound-field information with the visual modality to use such distal sensing for ASDL in an audio-visual system, including applications in immersive media production.

\section{ACKNOWLEDGMENT}
\label{sec:ack}

Research was funded by EPSRC-BBC Prosperity Partnership `{AI4ME}: Future personalised object-based media experiences delivered at scale anywhere' (EP/V038087/1), and a Doctoral College PhD studentship at the University of Surrey. 
For the purpose of open access, the authors have applied a Creative Commons Attribution (CC BY) license to any Author Accepted Manuscript version arising. 
Data supporting this study are available from \url{https://cvssp.org/data/TragicTalkers}.

\bibliographystyle{IEEEtran}
\bibliography{2023_WASPAA}
%
%
%
%
%
%
%
%
%

\end{sloppy}
\end{document}